\documentclass[10pt]{article}
\usepackage[preprint]{tmlr}
\setcitestyle{numbers,square,citesep={,}}

\usepackage{amsmath,amssymb,amsfonts}
\usepackage{graphicx}
\usepackage{booktabs}
\usepackage{array}
\usepackage{float}
\usepackage{hyperref}
\usepackage{url}
\DeclareUrlCommand\path{\urlstyle{tt}}
\usepackage{newunicodechar}
\newunicodechar{✓}{\ensuremath{\checkmark}}
\newunicodechar{✗}{\ensuremath{\times}}

\title{QuantumNovelty: A Skill-Orchestrating Language Agent for Referee-Style Review and Patentability Screening of Quantum Papers and Patents}

\author{\name Shlomo Kashani \email skashan2@alumni.jh.edu \\
      \addr Johns Hopkins University \\ QNeura.ai}

\begin{document}
\maketitle

\begin{abstract}
Language-model agents increasingly produce quantum-science results; we ask whether the same agentic paradigm can also scrutinize them in an auditable, reproducible, and cost-transparent form. We present QuantumNovelty, an open-source skill-orchestrating language agent that both generates quantum-computing artifacts (papers, Pareto-front ansatz candidates, and patent drafts) and reviews them through simulated referee and patent-examiner panels. Its design contribution is an audit-and-falsify layer of deterministic gates --- strict Pareto domination, numerical recomputation from on-disk artifacts, Wilson small-sample intervals, and a cross-vendor consensus guard --- that constrains, rather than generates, the claims allowed to survive; every model call is logged with backend, token count, and cost. We make no accuracy claim against human experts, and validate only what is checkable without human labels: on a planted adversarial corpus the deterministic gates catch every planted overclaim with no false positives, and on a first deployment (six manuscripts and one granted patent, at a measured cost of about twenty-four US dollars) the panels are directionally more conservative than the public acceptance record, on a one-sided sample. The framework is decision support, not a replacement for peer review or patent examination, and we report in full where its mechanisms remain unexercised on real inputs.
\end{abstract}

\begin{center}\small
Code: \url{https://github.com/BoltzmannEntropy/QuantumNovelty}\\
Project page: \url{https://boltzmannentropy.github.io/QuantumNovelty.github.io/}
\end{center}
\section{Introduction}

Quantum computing has entered a regime in which both the hardware and the literature scale faster than human expert attention. Processors have grown into the noisy intermediate-scale quantum era~\cite{preskill2018nisq} and beyond, with surface-code demonstrations on transmon-qubit~\cite{koch2007transmon} processors marking the path toward fault tolerance~\cite{googleqec2023}, while the variational and simulation algorithms that run on these devices~\cite{peruzzo2014vqe,cerezo2021vqa} generate a torrent of manuscripts and patents. Two pressures follow. The production of results is being automated, and so must their scrutiny.

Language-model agents, built on the transformer architecture~\cite{vaswani2017attention} and the in-context and chain-of-thought abilities of large models~\cite{brown2020gpt3,wei2022cot}, have moved from writing code to running science. They now act through tools and structured reasoning loops~\cite{yao2023react}, drive autonomous chemistry and materials laboratories~\cite{boiko2023coscientist,bran2023chemcrow}, and in quantum hardware bring up and calibrate a 112-qubit superconducting processor with little human intervention by orchestrating a library of reusable calibration skills~\cite{xu2026vibecalibration}. In parallel, program-search and discovery agents propose mathematical constructions, algorithms, and physical designs~\cite{romera2024funsearch,novikov2025alphaevolve,lu2024aiscientist}. As more of the production of quantum results is delegated to agents, the artifacts that document and scrutinize those results become a bottleneck of their own. A referee report, a prior-art analysis, a revision plan, and a cost ledger are themselves structured outputs that an agent can produce, and producing them in a uniform and auditable form has value independent of whether the underlying judgment matches a human's.

QuantumNovelty targets this production-and-scrutiny task for the quantum-computing literature and patent corpus. It is a skill-orchestrating language agent in the same architectural sense as recent hardware-bring-up systems~\cite{xu2026vibecalibration}, but it is pointed at the publication and patent plane rather than the control plane. The framework does not assert that it referees or examines as well as a human. It asserts something narrower and checkable. It produces referee-style and examiner-style artifacts through composable skills, it records the full provenance of every model call, and it refuses to emit certain unfalsifiable claims by construction.

We are careful about the motivation, because an earlier framing risked circularity. One could justify the framework by appealing to a ``novelty adjudication problem'' and then define that problem using the framework's own machinery. We avoid this. The motivation is concrete and prior to any design choice. Producing a paper review, a patent Office Action, or a novelty assessment today yields text and numbers that are neither uniform nor traceable. Two reviewers record their reasoning differently, numerical claims are restated by hand, prior-art citations are not machine-checkable, and the cost of producing a review is invisible. QuantumNovelty makes these artifacts uniform, recomputed, costed, and reproducible. Whether its judgments are good is a separate empirical question that this paper does not resolve. Concretely, we validate three claims, none of which requires human labels: (i) the deterministic audit gates behave as specified on planted inputs, catching every planted overclaim with no false positives --- a test of gate logic, not of extraction from real manuscripts (Section~\ref{sec:adversarial}); (ii) on the public outcome record the panels are directionally more conservative than the human disposition, on one-sided samples (Section~\ref{sec:experiments}); and (iii) every recorded model call writes a provenance sidecar, carrying token counts and monetary cost to the extent the backend reports them (Section~\ref{sec:provenance}). We do not claim that the framework's referee or examiner judgments match a human expert's; that would require the labeled evaluation of Section~\ref{sec:evalprotocol}, which we specify but, lacking annotator resources, do not run. The contribution is therefore a piece of auditable AI-for-science infrastructure with a bounded, checkable claim, not an assertion of human-level reviewing.

This paper describes the system and a first deployment. Section~\ref{sec:related} places it relative to agentic quantum systems. Section~\ref{sec:arch} describes the framework architecture and the full execution flow. Section~\ref{sec:modes} describes the two operating modes and the complete skill catalog. Section~\ref{sec:audit} details the audit-and-falsify layer. Section~\ref{sec:pipeline} walks through the full generative pipeline. Section~\ref{sec:provenance} describes provenance. Section~\ref{sec:review} reports review-mode case studies on real manuscripts with quoted findings. Section~\ref{sec:patent} is a dedicated patent-examination case study. Section~\ref{sec:validation} describes implementation validation. Section~\ref{sec:limitations} discusses limitations.

\section{Related Work}
\label{sec:related}

\subsection{Skill-orchestrating language agents for quantum science}

The closest system in spirit is Vibe Calibration, which brings up a 112-qubit superconducting quantum processor through a language agent that orchestrates a library of structured calibration skills~\cite{xu2026vibecalibration}. Each skill in that system is a parameterized decision tree that encodes a measurement protocol, a quantitative acceptance criterion, and rollback logic for failure recovery. The agent grows the library through a three-phase curriculum that moves from human-guided to semi-autonomous to fully autonomous operation. The authors report calibrating 108 of 112 qubits autonomously in 4.7 hours, a four- to five-fold speedup over manual calibration, with agent-derived coherence metrics that are statistically indistinguishable from an expert's on a controlled 16-qubit comparison~\cite{xu2026vibecalibration}.

QuantumNovelty shares the architecture and occupies an orthogonal quadrant of the same design space. Vibe Calibration sits at the intersection of experiment and hardware, and its governance artifact is a calibration audit log. QuantumNovelty sits at the intersection of publication and literature, and its governance artifacts are a referee-style report, a simulated patent Office Action, and a provenance ledger. The two systems together suggest a single agent paradigm spanning both the production and the documentation of quantum results.

\subsection{Agentic research-automation systems}
\label{sec:priorart}

QuantumNovelty is best understood as a domain specialization of a broader class of general-purpose agentic research systems, and it inherits concrete machinery from two of them. AutoResearchClaw (ARC) is an autonomous research pipeline that takes a topic to a full paper through a chain of staged skills, with deterministic gates for citation integrity, a quality gate, a research-decision verdict, and a knowledge archive~\cite{arc_autoresearchclaw}. Academic Research Skills (ARS) is a suite of Claude Code skills for academic research, paper authoring, multi-voice peer review, and pipeline orchestration, including a thirteen-agent deep-research skill~\cite{ars_academicskills}. QuantumNovelty borrows directly from both: its skill-as-folder layout and its paper-authoring and reviewer skills follow the ARS pattern, while its deterministic zero-model gates, the claims registry and the evidence ledger, are adapted from ARC's verification stages. Hermes is a general personal agent that runs one agent core across a command-line interface, a messaging gateway, a terminal user interface, and a desktop application, learning across sessions through memory and skills and delegating to subagents~\cite{hermes_agent}. Where ARC and ARS automate the general research-and-writing loop and Hermes provides a general agent substrate, QuantumNovelty narrows the target to quantum computing and adds two things those systems do not provide: a strict-domination audit-and-falsify layer tuned to quantum-circuit objectives, and a six-voice patent-examination subsystem keyed to United States patent law. The relationship is therefore one of specialization and extension rather than competition. The same agent paradigm appears in general LLM-as-judge evaluation~\cite{zheng2023judge} and in studies of large-scale automated reviewer feedback on scientific manuscripts~\cite{liang2024feedback}, which QuantumNovelty specializes to the quantum domain with an explicit, auditable rubric.

To ground this section we did not curate the prior art by hand. We ran QuantumNovelty's own \texttt{literature\_surfacer} skill on this paper's topic, querying Crossref, arXiv, and Semantic Scholar, and let the framework surface its own related work. The relevant returns include systematic reviews of large-language-model-powered automated assessment~\cite{emirtekin2025assessment}, retrieval-augmented agents for automated scientific literature review~\cite{wang2026ragreview}, programming languages for orchestrating agents~\cite{mohammadi2025pel}, large-language-model novelty detection~\cite{seifert2026novelty}, and the patent-examination practice of patentability, or novelty, search that QuantumNovelty's patent subsystem automates~\cite{le2026patentability}. This is the framework reviewing its own manuscript's prior art. Consistent with the limitations in Section~\ref{sec:limitations}, the broad query also returned off-topic hits, for example agents for building-energy management and smart-grid defense, that a human still has to filter. Automated surfacing widens recall but does not replace editorial judgment, which is why prior-art search in QuantumNovelty is a stage whose output is logged and auditable rather than an unquestioned oracle.

\subsection{Automated discovery and the novelty problem}

A growing family of agents searches program or design spaces and reports discoveries, including mathematical constructions from program search~\cite{romera2024funsearch}, broad algorithmic and scientific discovery~\cite{novikov2025alphaevolve}, and end-to-end automated research~\cite{lu2024aiscientist}. These systems motivate the audit-and-falsify layer directly. A reported improvement may be a rediscovery of known prior art, an interpolation between cataloged results, or a genuine strict domination of the baseline. Distinguishing these cases requires an explicit baseline catalog and a domination test. The quantum manuscripts in our deployment corpus exemplify the targets of such scrutiny, spanning variational eigensolvers~\cite{peruzzo2014vqe, cerezo2021vqa}, Hamiltonian-simulation error compensation~\cite{zeng2025lcutrotter}, generative warm-starts for the variational quantum eigensolver~\cite{zou2025flowvqe}, quantum convolutional networks~\cite{cong2019qcnn}, and trainability questions such as barren plateaus~\cite{mcclean2018barren}, all within the noisy intermediate-scale quantum regime~\cite{preskill2018nisq}. The complementary failure axis, the logical correctness of an agent-produced argument, is addressed by machine verification: a recent autoformalization loop produced a Lean-verified proof of a decade-old quantum-optimization conjecture, a bounded world model in which a hallucinated proof step cannot survive compilation~\cite{kol2026machineverified}. That approach certifies that a result is right; it does not ask whether the underlying insight was already known. QuantumNovelty targets precisely the axes formal verification leaves open, rediscovery, interpolation, and empirical overclaim, and the two verification styles are complementary rather than competing. In the same spirit, scientific claim-verification corpora~\cite{wadden2020scifact,wadden2022multivers} are the natural-language analogue of the review mode's fallacy detection, and quantum-circuit benchmark suites~\cite{li2023qasmbench,quetschlich2023mqtbench} provide the reference corpora against which a literature-populated Pareto catalog would be measured.

\subsection{Deterministic verification infrastructure}

QuantumNovelty's deterministic skills draw on established scholarly infrastructure. Citation verification queries Crossref metadata~\cite{m2024crossref} with no language model in the loop. Small-sample rate reporting uses the Wilson score interval~\cite{wilson1927probable}. The framework's refusal to accept unfalsifiable claims is an operational reading of a falsificationist standard for scientific assertions~\cite{popper1959logic}. Several deterministic gates are adapted from two upstream agent toolkits for research automation and skill composition~\cite{arc_autoresearchclaw, ars_academicskills}.

\subsection{Positioning against LLM-as-judge and automated review}
\label{sec:positioning}

Table~\ref{tab:related} places QuantumNovelty against the closest cited systems along the axes that matter for an auditable review tool: whether outputs are checked against any external standard, whether the pipeline contains model-free deterministic gates, and whether every model call is logged with its cost. General LLM-as-judge evaluation and large-scale automated-reviewer-feedback studies~\cite{zheng2023judge, liang2024feedback} assess against human agreement but carry no deterministic falsification gates and no per-call provenance ledger. Agentic research systems~\cite{lu2024aiscientist, arc_autoresearchclaw, ars_academicskills} automate the writing-and-review loop but are not pointed at the quantum-review-and-patent plane and report no external calibration. Vibe Calibration~\cite{xu2026vibecalibration} shares the skill-orchestration architecture but governs hardware bring-up rather than the publication plane. QuantumNovelty's distinguishing combination is a domain-specific deterministic audit layer (strict Pareto domination, numerical recomputation, Wilson intervals, and a cross-vendor guard), a full per-model-call provenance ledger, and an evaluation that pairs coarse outcome-level comparison (journal acceptance, patent grant) with the adversarial gate test of Section~\ref{sec:adversarial}. We claim differentiation on this combination, not on any single mechanism, several of which are inherited from the upstream toolkits above.

\begin{table}[t]
\centering
\caption{QuantumNovelty against closely related agentic and evaluation systems, grouped by class. ``Det.\ gates'' marks model-free deterministic checks in the pipeline; ``Provenance'' marks per-model-call logging of backend, tokens, and cost; ``Cost'' is the measured per-run monetary cost reported \emph{in this paper}; ``\textemdash'' marks a system whose paper does not report a dollar figure, which is not evidence that it lacks cost instrumentation.}
\label{tab:related}
\footnotesize
\setlength{\tabcolsep}{5pt}
\begin{tabular}{@{}p{2.2cm}p{2.4cm}p{2.1cm}p{3.1cm}p{1.9cm}p{1.8cm}@{}}
\toprule
\textbf{System} & \textbf{Task} & \textbf{Domain} & \textbf{Det.\ gates} & \textbf{Provenance} & \textbf{Cost} \\
\midrule
\multicolumn{6}{@{}l@{}}{\emph{General LLM evaluation and review}} \\
\quad LLM-as-judge~\cite{zheng2023judge} & Pairwise eval & General & No & No & \textemdash \\
\quad Reviewer feedback~\cite{liang2024feedback} & Review comments & General science & No & No & \textemdash \\
\midrule
\multicolumn{6}{@{}l@{}}{\emph{Agentic research automation}} \\
\quad AI Scientist~\cite{lu2024aiscientist} & End-to-end research & General ML & No & Partial & \textemdash \\
\quad ARC / ARS~\cite{arc_autoresearchclaw, ars_academicskills} & Research \& writing & General academic & Citation, quality & Partial & \textemdash \\
\midrule
\multicolumn{6}{@{}l@{}}{\emph{Skill-orchestrating quantum agents}} \\
\quad Vibe Calib.~\cite{xu2026vibecalibration} & HW calibration & Quantum HW & Acceptance criteria & Calib.\ log & \textemdash \\
\midrule
\textbf{This work} & Review + patent screening & Quantum + patents & Pareto, recompute, Wilson & Full per-call & \$24.51$^{\star}$ \\
\bottomrule
\end{tabular}\\[3pt]
{\footnotesize\raggedright $^{\star}$Total measured cost of the deployment of Section~\ref{sec:review} (86 model calls, roughly \$3 per referee-panel run). The dashes mark values the compared systems' papers do not report, not an absence of cost instrumentation.}
\end{table}

\section{Framework Architecture}
\label{sec:arch}

\begin{figure}[t!]
  \centering
  \includegraphics[width=0.92\textwidth]{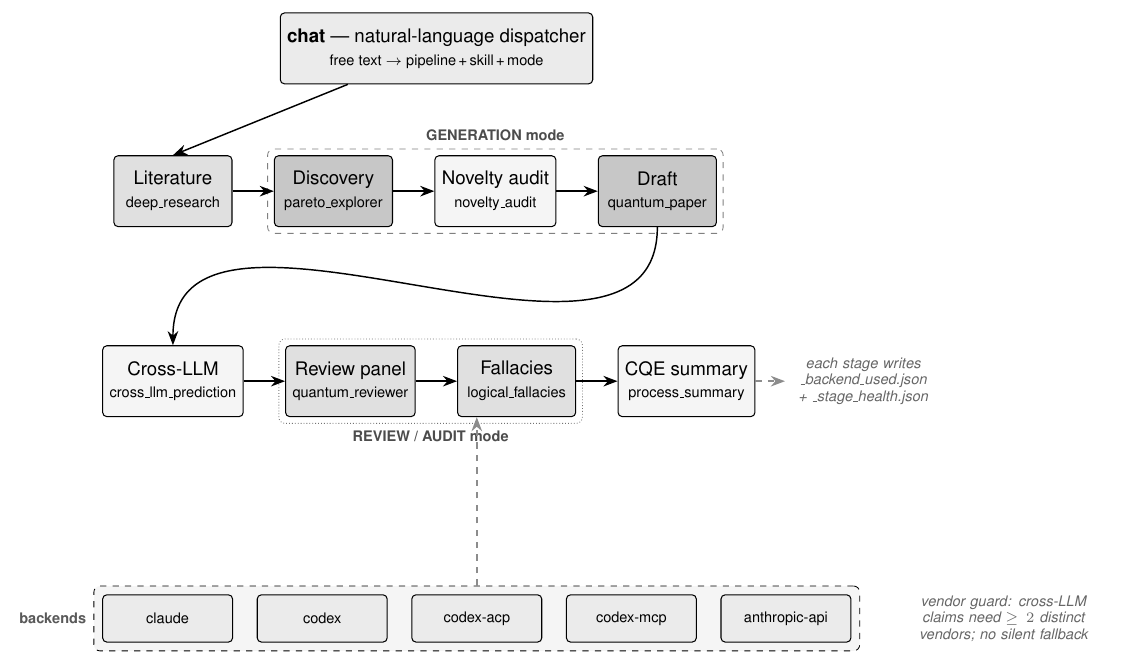}
  \caption{QuantumNovelty architecture and the full generative pipeline. A
  natural-language dispatcher routes a free-text request to a pipeline, skill,
  and mode. The full pipeline chains eight stages, shaded by operating
  mode (darkest for generation, mid for review and audit, lightest for
  deterministic stages), with dashed and dotted bands labelling the
  generation and review/audit groups. Each stage writes provenance and stage-health
  sidecars. All stages run on a substrate of five backends behind one call
  interface, guarded by a cross-model vendor rule and a no-silent-fallback
  policy.}
  \label{fig:flow}
\end{figure}

QuantumNovelty is structured as a set of skills, a shared library, a chain dispatcher, and a natural-language front end (Fig.~\ref{fig:flow}). The architecture chains a natural-language dispatcher that routes a free-text request to a pipeline, skill, and mode; a full pipeline of eight stages grouped by operating mode (generation, review and audit, and deterministic stages); per-stage provenance and stage-health sidecars; and a substrate of five backends behind one call interface, guarded by a cross-model vendor rule and a no-silent-fallback policy.

\subsection{Skill-as-folder pattern}

Each skill is a self-contained directory that contains a machine-readable description, a shell entry point, and a Python driver. A skill declares its inputs and outputs in its header, and the chain composes skills by name. The framework defines twenty-two skill directories in addition to the shared common module. The skills fall into paper review, patent examination and drafting, paper and ansatz generation, deterministic verification, literature acquisition, and infrastructure.

\subsection{Common library and backends}

A shared module provides one calling convention across every backend. The framework declares exactly five backends behind a single call interface. The default backend shells out to the Claude Code command-line interface, which means the default path requires no vendor API key and bills against an existing subscription. A second backend shells out to a Codex command-line tool, which serves as a distinct vendor for cross-model falsification. Two further backends route Codex through the Agent Client Protocol for a persistent named session across skill calls, so a later stage can use context built by an earlier one, and through a Model Context Protocol~\cite{mcp2024} server that exposes QuantumNovelty's own skills as tools. A fifth backend calls a vendor HTTP API directly.

Two policies harden the backend layer. First, there is no silent fallback. A backend whose binary is missing raises an error rather than downgrading to another vendor, and an unknown backend name is rejected at the call boundary. Second, the default backend pins a specific model identifier on each call, which prevents the command-line interface from adaptively downgrading a short prompt to a smaller model. The provenance ledger in Section~\ref{sec:provenance} confirms that real runs used three distinct model identifiers and recorded which one served each call.

\subsection{The full execution flow}

A pipeline executes as an ordered loop. The shell entry point parses the pipeline name, the backend, the output root, and the checkpoint-control flags, then delegates to a Python stage executor. For each stage the executor first checks whether the stage is already complete or is being skipped by a resume request. It then creates a numbered output subdirectory, writes a stage-health record marked running, and invokes the skill as a subprocess with the stage subdirectory and the backend passed through. On return it updates the stage-health record to done or failed, appends a decision entry to the run's decision history, and records a per-stage result holding the stage name, the skill, the return code, and the elapsed time. If a pause-after flag matches the current stage, the executor writes a checkpoint and exits cleanly so a human can inspect the partial run before resuming.

The decision history records two runtime verdicts, proceed when a skill exits zero and fail when it does not. The final disposition of a run is proceed when every stage returned zero and fail otherwise. This is a deliberately simple control model. The framework does not attempt automatic refine-or-pivot branching at the chain level. It records what happened and leaves the decision to re-run, resume, or abandon to the operator, supported by the checkpoint controls. Three flags govern non-linear execution. A pause-after flag stops cleanly after a named stage, a resume-from flag treats earlier stages as complete, and a force flag re-runs stages whose output already exists.

Every pipeline ends with a terminating summary skill that reads every prior stage's outputs, reconstructs the process timeline, and scores the run on six dimensions of collaboration quality on a zero-to-one-hundred scale. The six dimensions are combined with a geometric mean rather than an arithmetic one, so a zero on any single dimension collapses the composite. This cost-quality-evidence summary is the framework's self-assessment of a run; as Section~\ref{sec:limitations} discusses, in the deployment reported here it registered the framework's own unfilled capabilities rather than discriminating paper quality.

\subsection{The natural-language dispatcher}

A chat skill maps a free-text request to a pipeline, a skill, a mode, and flags, so a user can type ``Review this paper'' instead of recalling the exact command. The five named pipelines it can route to are \texttt{novelty-audit} (the audit-and-falsify workflow), \texttt{full} (the legacy end-to-end ansatz-discovery workflow), \texttt{paper-audit} (review of an existing paper), \texttt{patent-audit} (the six-voice examiner panel), and \texttt{scout} (topic-driven novelty scouting). The dispatcher matches patterns first, which is cheap and deterministic, and falls back to model classification when no pattern matches. It writes a structured dispatch decision recording the chosen skill, mode, flags, and a confidence value, and it executes the dispatched skill only when explicitly asked to. Representative routings send ``Write a paper on X'' to the authoring skill in full mode, ``Review this paper'' to the reviewer skill in full mode, ``Find fallacies in this paper'' to the fallacy skill, and ``I already have a paper, review it'' to a mid-entry review pipeline.

\section{Two Operating Modes}
\label{sec:modes}

QuantumNovelty's skills divide into a generation mode and a review-and-audit mode. We describe both, and we are explicit in Section~\ref{sec:limitations} about which has been exercised end to end on real inputs.

\subsection{Generation mode}

Generation mode authors new artifacts. The paper-authoring skill offers ten modes spanning the lifecycle of a manuscript, summarized in Table~\ref{tab:authoring}. It is venue-aware, in that every mode reads the target journal's policy, and library-aware, in that the code-generating modes emit working snippets in the user's chosen quantum library.

\begin{table}[H]
\centering
\caption{The ten modes of the paper-authoring skill.}
\label{tab:authoring}
\footnotesize
\begin{tabular}{l p{0.62\columnwidth}}
\toprule
\textbf{Mode} & \textbf{Purpose} \\
\midrule
full & Write a complete first draft from a topic \\
plan & Guided contribution-then-method-then-results planning \\
outline-only & Section outline with per-section word budgets keyed to the venue \\
revision & Apply reviewer comments, producing a revised draft and a response letter \\
revision-coach & Parse reviewer comments into a roadmap without applying them \\
abstract-only & Write only the abstract, sized to the venue limit \\
lit-review & Rewrite as a literature-review paper \\
format-convert & Convert between templates and citation styles \\
citation-check & Verify every citation resolves and matches its claim \\
disclosure & Generate the AI, data, code, conflict, and funding disclosure block \\
\bottomrule
\end{tabular}
\end{table}

The ansatz-discovery skill drives a language-model mutation loop over quantum-circuit ans\"atze and builds a strict-domination Pareto archive over energy error, parameter count, gate count, and two-qubit-gate count. Each generation the model proposes candidate circuits, an evaluator measures them against a fixed Hamiltonian, and non-dominated points join the archive. It offers a built-in evaluator using a bundled state-vector simulator with a registry of transverse-field Ising, Heisenberg, and hydrogen-molecule Hamiltonians, an external-evaluator mode that shells out to a user command, and a plan-only mode that writes the run plan without any model or evaluator call. The patent-drafting skill starts from an invention disclosure and produces a full filing package, and an ablation-design skill builds the controls that distinguish a real model contribution from a lucky random walk, with axes that toggle the model mutator, the commutation hint, Pareto seeding, and a cross-vendor swap.

\subsection{Review and audit mode}

Review mode scrutinizes existing artifacts. The paper-reviewer skill offers six modes, summarized in Table~\ref{tab:reviewer}. The research skill offers seven modes, from a full multi-source literature pull suitable for novelty-audit augmentation, through PRISMA-style systematic review, a Socratic question-formulation dialogue, targeted fact-checking, and a research-quality review of a candidate paper. The patent-reviewer skill offers a full six-voice screening report and a quick single-voice triage.

\begin{table}[H]
\centering
\caption{The six modes of the paper-reviewer skill.}
\label{tab:reviewer}
\footnotesize
\begin{tabular}{l p{0.62\columnwidth}}
\toprule
\textbf{Mode} & \textbf{Purpose} \\
\midrule
full & Complete five-voice panel: Editor-in-Chief, three reviewers, Devil's Advocate \\
quick & One-page assessment by a single reviewer \\
guided & Iterative coaching that surfaces weaknesses and suggests revisions \\
methodology-focus & Panel restricted to methodology and audit-framework adherence \\
re-review & Verify a revised draft addressed the prior round's comments \\
calibration & Run against a gold set of known-good and known-flawed papers and report reliability \\
\bottomrule
\end{tabular}
\end{table}

The remaining review-mode skills are the audit-and-falsify layer of Section~\ref{sec:audit} together with supporting deterministic skills. Table~\ref{tab:catalog} is the full skill catalog grouped by function.

\begin{table}[t!]
\centering
\caption{The full skill catalog grouped by function.}
\label{tab:catalog}
\footnotesize
\begin{tabular}{p{0.16\textwidth} l p{0.58\textwidth}}
\toprule
\textbf{Group} & \textbf{Skill} & \textbf{One-line role} \\
\midrule
Paper review & quantum\_reviewer & Five-voice referee panel, six modes \\
Paper review & deep\_research & Topic research and literature synthesis, seven modes \\
Paper review & logical\_fallacies & Named-fallacy detection with a quantum-specific taxonomy \\
Paper review & argument\_structure & Premise-to-conclusion map, claim-proof gap, narrative-debt register \\
Paper review & revision\_planner & Paragraph-anchored, verbatim-quoted revision roadmap \\
Paper review & disclosure\_audit & Sixteen-point disclosure-completeness checklist \\
Paper review & requirements\_judge & Claim-versus-evidence audit against an allowed-or-forbidden manifest \\
Audit and falsify & novelty\_audit & Strict Pareto domination, ratio recompute, Wilson intervals, honest negatives \\
Audit and falsify & audit\_falsify & Low-level domination, classification, and interval primitives \\
Audit and falsify & cross\_llm\_prediction & Same prompt across distinct vendors with the vendor guard \\
Deterministic gates & citation\_integrity & Four-layer Crossref citation verifier, zero model calls \\
Deterministic gates & claims\_registry & Deterministic numeric-claim audit, zero model calls \\
Deterministic gates & evidence\_ledger & Pre-registers paper facts, then flags unanchored attributions \\
Patent & patent\_reviewer & Six-voice USPTO examiner panel \\
Patent & patent\_drafter & Full filing-package drafting \\
Generation & quantum\_paper & Quantum-paper authoring, ten modes \\
Generation & pareto\_explorer & Pareto-front ansatz discovery loop \\
Generation & ablation\_designer & Designs and runs ablation controls \\
Literature & literature\_surfacer & Crossref, arXiv, and Semantic Scholar pull \\
Literature & book\_acquirer & Book and thesis acquisition with OCR \\
Infrastructure & process\_summary & Terminating cost-quality-evidence summary \\
Infrastructure & chat & Natural-language dispatcher \\
\bottomrule
\end{tabular}
\end{table}

\section{The Audit-and-Falsify Layer}
\label{sec:audit}

\begin{figure}[t]
  \centering
  \includegraphics[width=\columnwidth]{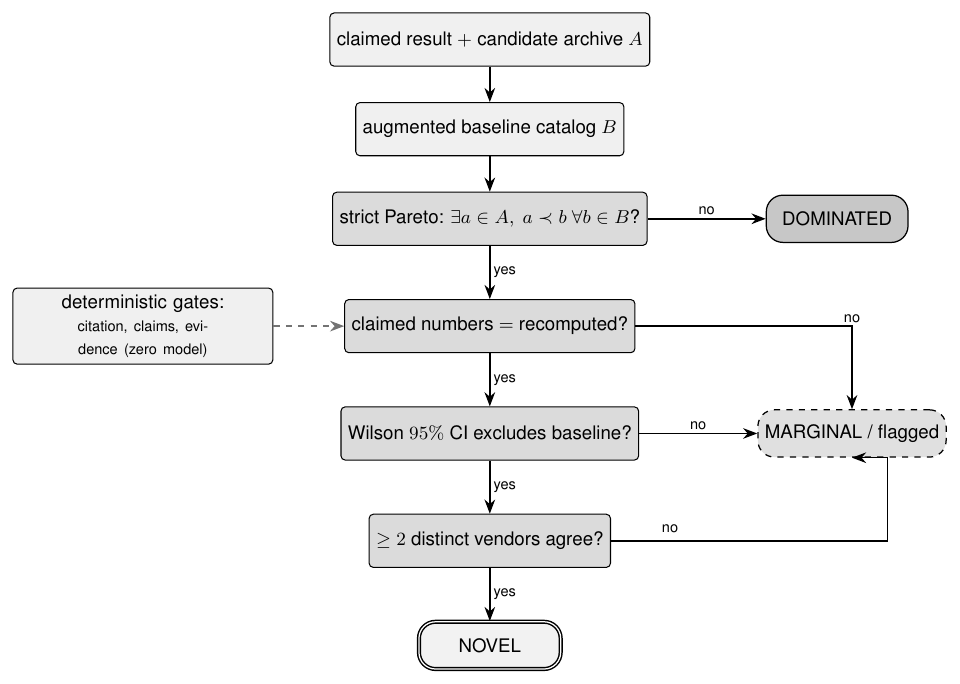}
  \caption{The audit-and-falsify decision flow. A claimed result is certified
  \textsc{novel} only if it strictly Pareto-dominates the augmented baseline,
  its numbers survive recomputation, its small-sample rate clears a Wilson
  interval, and at least two distinct vendors agree. Any failed gate routes to
  \textsc{dominated} or \textsc{marginal}. Three deterministic, zero-model
  gates (citation, claims, evidence) run alongside.}
  \label{fig:audit}
\end{figure}

The audit-and-falsify layer encodes a reviewer's skepticism as a sequence of explicit, mostly deterministic checks, each with a stated acceptance criterion. The design principle is that the burden of proof rests on the claim, not on the reviewer: a result is escalated from candidate to accepted only if it survives every check in turn. Throughout, ``accepted'' (and, in the figures, ``certified'') denotes only that a claim passed the framework's deterministic gates on the evidence supplied; it is not a determination that the claim is scientifically true, genuinely novel, or patentable, all of which require external ground truth this deployment does not have. Figure~\ref{fig:audit} shows the decision flow, and Table~\ref{tab:criteria} is the single source of truth for every gate criterion. We describe each mechanism, give its formal criterion, and state whether the deployment corpus actually exercised it, because an implemented mechanism and an exercised mechanism are not the same thing.

\subsection{Formal criteria}

\paragraph{Strict Pareto domination.} A novelty claim about a quantum-circuit ansatz is evaluated against an augmented catalog $B$ of baseline results rather than a single reference, on a vector of objectives $\mathbf{f}(\cdot)=(\Delta E,\, n_p,\, n_g,\, n_{2q})$, namely energy error, parameter count, gate count, and two-qubit-gate count, in which lower is better on every axis. A candidate $a$ in the discovery archive $A$ strictly dominates a baseline $b$, written $a\prec b$, when
\begin{equation}
a \prec b \iff f_i(a)\le f_i(b)\ \forall i \ \wedge\ \exists j:\, f_j(a) < f_j(b).
\end{equation}
The verdict is \textsc{novel} only if the catalog is non-empty and some candidate dominates all of it, $B\neq\emptyset \,\wedge\, \exists a\in A:\ a\prec b\ \forall b\in B$; the non-emptiness conjunct is explicit so that an empty catalog can never yield \textsc{novel} by vacuous quantification. Otherwise the result is \textsc{dominated} or, if it improves one axis while regressing another, \textsc{marginal}. The catalog is augmented (seeded with strong known results) so domination is not won against a deliberately weak field. This mechanism requires a populated archive, which, as Section~\ref{sec:limitations} notes, the literature corpus did not supply.

\paragraph{Numeric recomputation.} Every quantitative claim routed through the recomputation gate --- displayed ratios and small-sample success rates --- is checked against on-disk artifacts before acceptance; a claimed value $r_{\text{claim}}$ is accepted only when it matches the value $r_{\text{recomp}}$ derived from stored run artifacts within a tolerance $\tau$,
\begin{equation}
\lvert r_{\text{claim}} - r_{\text{recomp}}\rvert \le \tau,
\end{equation}
and is flagged otherwise; in the reported runs $\tau$ is a relative tolerance of two percent, i.e.\ the accepted band is $\lvert r_{\text{claim}} - r_{\text{recomp}}\rvert \le 0.02\,\lvert r_{\text{recomp}}\rvert$. The layer never trusts a model's restatement of a number it produced earlier in the same run. The patent screening report of Section~\ref{sec:patent} instantiates this: its structured record is a deterministic parse of the prose rather than a second model call, so the per-claim rejection is a fixed function of the panel's stated reasoning rather than an independent model judgment (though, as Section~\ref{sec:patent} shows, a defect in that parse can still misread the prose).

\paragraph{Wilson small-sample intervals.} Agent evaluations report success over few trials, so a bare rate is misleading. For $k$ successes in $n$ trials the layer reports the Wilson $95\%$ score interval~\cite{wilson1927probable},
\begin{equation}
\frac{\hat p + \dfrac{z^2}{2n} \pm z\sqrt{\dfrac{\hat p(1-\hat p)}{n} + \dfrac{z^2}{4n^2}}}{1 + \dfrac{z^2}{n}},\qquad \hat p=\frac{k}{n},\ z=1.96,
\end{equation}
and a method is reported to beat a baseline only when this interval excludes the baseline rate. This makes three-of-four trials and thirty-of-forty trials read as the different evidentiary claims they are.

\paragraph{Cross-model vendor guard.} A claim resting on agreement among models is accepted only when the agreement spans distinct commercial vendors. A vendor-resolution function $v(\cdot)$ maps each backend name to a vendor (for example any \texttt{claude*} name to one vendor and any \texttt{codex*}/\texttt{gpt*} name to another). For a set of backends $L$ the guard requires
\begin{equation}
\bigl\lvert \{\, v(b) : b\in L \,\}\bigr\rvert \ge 2,
\end{equation}
and the skill exits with a nonzero status otherwise, so two snapshots of one model family never count as cross-model evidence. The recorded vendor set travels with the prediction, making the multi-vendor property checkable from the artifact alone.

\paragraph{Panel and composite thresholds.} The five-voice review panel accepts a manuscript only when the mean voice score clears a fixed bar, $\bar s \ge 7.0$ out of $10$. The terminating cost-quality-evidence (CQE) summary scores a run on six dimensions $d_1,\dots,d_6 \in [0,100]$ and combines them with a geometric mean,
\begin{equation}
C = \Bigl(\textstyle\prod_{i=1}^{6} d_i\Bigr)^{1/6},
\end{equation}
so that a zero on any single dimension collapses the composite to zero. This deliberately punishes a run that is strong on most axes but absent on one (for example a paper with no reproducibility artifacts), rather than letting an arithmetic mean average the gap away.

\paragraph{Deterministic zero-model gates.} Three gates run with no model call and are therefore reproducible across runs and machines. A citation-integrity gate verifies references against Crossref~\cite{m2024crossref} through a four-layer check with no retrieval-augmented generation. A claims-registry gate performs numeric-claim audits deterministically. An evidence-ledger gate pre-registers the facts a paper actually states and then scans every review report for an attribution the paper never made, closing a hole the review panel cannot close on its own.

\begin{table}[t]
\centering
\caption{Acceptance criteria for every audit-and-falsify check (the framework's single source of truth). Deterministic gates make no model call and are reproducible across runs.}
\label{tab:criteria}
\footnotesize
\begin{tabular}{p{0.20\textwidth} p{0.34\textwidth} p{0.38\textwidth}}
\toprule
\textbf{Check} & \textbf{What it tests} & \textbf{Gate criterion} \\
\midrule
Strict Pareto domination & candidate ansatz vs augmented baseline on (energy error, params, gate count, two-qubit gates) &\textsc{novel} iff $\exists a\in A:\ a\prec b\ \forall b\in B$; else \textsc{dominated}/\textsc{marginal} \\
Numeric recomputation & claimed numbers vs on-disk artifacts & accept iff $\lvert r_{\text{claim}}-r_{\text{recomp}}\rvert\le\tau$; else flag \\
Wilson interval & small-sample success rate & report $95\%$ Wilson interval; ``beats baseline'' only if interval excludes baseline \\
Cross-model vendor guard & multi-model consensus & proceed iff $\lvert\{v(b)\}\rvert\ge 2$ distinct vendors; else exit nonzero \\
Panel pass & five-voice mean score & accept iff $\bar s\ge 7.0/10$ \\
CQE composite & six quality dimensions & $C=(\prod_{i=1}^6 d_i)^{1/6}$; any zero collapses $C$ \\
Citation integrity (det.) & references vs Crossref & four-layer match, zero model calls \\
Claims registry (det.) & numeric claims in a document & deterministic recompute, zero model calls \\
Evidence ledger (det.) & attributions in review reports & flag any claim attributed to the paper it never made \\
\bottomrule
\end{tabular}
\end{table}

The CQE dimensions are themselves fixed and keyword-probed, listed in Table~\ref{tab:cqe}; Section~\ref{sec:limitations} discusses the consequence that all five external papers scored an identical composite.

\begin{table}[t]
\centering
\caption{The six cost-quality-evidence (CQE) dimensions, each scored $0$--$100$ and combined by geometric mean.}
\label{tab:cqe}
\footnotesize
\begin{tabular}{l p{0.60\columnwidth}}
\toprule
\textbf{Dimension} & \textbf{What it probes} \\
\midrule
Novelty rigor & augmented baseline catalog and strict-domination verdict present \\
Reproducibility & populated Pareto archive and stored run artifacts \\
Methodological rigor & ablations, statistics, and precision disclosures \\
Falsifiability & multi-vendor evidence and refutable claims \\
Domain depth & quantum-specific correctness and prior-art coverage \\
Communication & structure, clarity, and disclosure completeness \\
\bottomrule
\end{tabular}
\end{table}

\section{The Full Generative Pipeline}
\label{sec:pipeline}

The full pipeline chains the generation and review skills end to end, detailed in Table~\ref{tab:pipeline}. It begins with a literature stage that pulls sources and emits a Pareto-shaped baseline catalog. A discovery stage then runs the ansatz loop and writes an archive. A novelty-audit stage compares the archive against the baseline catalog and emits a verdict. A draft stage authors a paper if none was supplied. A cross-model stage forks an amplitude-prediction task across vendors. A review stage runs the five-voice panel, a fallacy stage scans the draft, and the terminating summary stage scores the whole run.

\begin{table}[t!]
\centering
\caption{The eight-stage full generative pipeline.}
\label{tab:pipeline}
\footnotesize
\begin{tabular}{c l l p{0.55\textwidth}}
\toprule
\textbf{\#} & \textbf{Stage} & \textbf{Skill} & \textbf{Produces} \\
\midrule
1 & Literature & deep\_research & Synthesis and a baseline catalog \\
2 & Discovery & pareto\_explorer & A strict-domination archive of candidates \\
3 & Novelty audit & novelty\_audit & A novel, dominated, or marginal verdict \\
4 & Draft & quantum\_paper & A paper draft, if none supplied \\
4a & Cross-model & cross\_llm\_prediction & Per-vendor predictions and a summary \\
5 & Review & quantum\_reviewer & A five-voice panel report \\
5b & Fallacies & logical\_fallacies & Named-fallacy findings \\
6 & Summary & process\_summary & The cost-quality-evidence scorecard \\
\bottomrule
\end{tabular}
\end{table}

The discovery, novelty-audit, and cross-model stages are conditional on the presence of a Hamiltonian or a prediction task. When a user supplies an existing paper, the framework offers mid-entry pipelines that skip the generation stages and run only the review and audit half, which is the path the deployment of Section~\ref{sec:review} used.

\section{Reproducibility and Provenance}
\label{sec:provenance}

Reproducibility is treated as a first-class output. Each stage gets a numbered subdirectory holding its outputs together with two sidecars. One records the backend requested, the backend actually used, the model identifier, input and output token counts, cache statistics, elapsed time, and monetary cost. The second records stage health in a fixed schema. The run as a whole writes a subprocess summary with the exit code and elapsed time of each stage, a telemetry aggregate that rolls up the per-stage health files, and a decision history of proceed and fail entries.

Two properties make the ledger trustworthy. First, the backend marker is written from the call site after every model call rather than inferred later, and a fidelity check compares the actually-used field against the requested field so that a silent routing change is caught rather than assumed impossible. Second, token counts come from the structured output envelope of the command-line interface when available, and the framework marks counts as estimated when it must fall back to a character-count proxy. In the deployment corpus, ten of eighty-six recorded calls carried estimated rather than measured token counts, and sixty-seven of eighty-six carried a measured cost; the nineteen calls without a cost figure are older sidecars that predate the cost-tracking field rather than failed calls, and the ledger flags exactly which ones.

\section{Review-Mode Case Studies}
\label{sec:review}

We report what the review mode actually produced on a corpus of six quantum-computing manuscripts, reviewed in eight referee-panel runs because two of the papers were run under both vendors. To fix the count once: the six comprise five external manuscripts (Flow-VQE, LCU-Trotter, HWQML, QCNN, Majorana) plus one internal manuscript; the panel table (Table~\ref{tab:panel}) lists all six, while the cost-quality-evidence tables of Section~\ref{sec:experiments} cover only the five external papers, since the internal one was not part of that scoring run. This is a usage demonstration, not an accuracy study, because the corpus carries no human-referee labels. We report what the framework computed, what it cost, and where its outputs diverged.

\subsection{Corpus and cost}

The corpus comprises paper-audit runs on Flow-VQE~\cite{zou2025flowvqe} (\href{https://arxiv.org/abs/2507.01726}{arXiv:2507.01726}), LCU-Trotter~\cite{zeng2025lcutrotter} (\href{https://arxiv.org/abs/2212.04566}{arXiv:2212.04566}, published as PRX Quantum \textbf{6}, 010359), a hardware-efficient quantum machine learning manuscript, a quantum convolutional neural network manuscript, a Majorana manuscript, and an internal manuscript, together with a patent-mode run whose calls enter the repository ledger; the patent case study of Section~\ref{sec:patent} was run separately against a granted patent and reports its own recorded cost. Across the repository the framework recorded eighty-six model-call sidecars, of which seventy-seven carried usage data. Table~\ref{tab:usage} aggregates the recorded usage, and Table~\ref{tab:sidecars} reconciles the overlapping sidecar subsets that the provenance discussion of Section~\ref{sec:provenance} reports individually, since the usage and cost partitions are independent and a reader cannot otherwise recover one from the other.

\begin{table}[H]
\centering
\caption{Recorded model-call usage aggregated across the deployment corpus.}
\label{tab:usage}
\begin{tabular}{lr}
\toprule
\textbf{Metric} & \textbf{Recorded total} \\
\midrule
Input tokens & 500{,}113 \\
Output tokens & 211{,}142 \\
Cache-read input tokens & 402{,}378 \\
Total cost (US dollars) & 24.51 \\
Total elapsed time (seconds) & 6{,}685.6 \\
\bottomrule
\end{tabular}
\end{table}

\begin{table}[H]
\centering
\caption{Sidecar accounting: overlapping subsets of the 86 recorded model calls. The usage and cost partitions are independent; a sidecar can carry usage data yet predate the cost-tracking field.}
\label{tab:sidecars}
\footnotesize
\begin{tabular}{lrl}
\toprule
\textbf{Sidecar subset} & \textbf{Count} & \textbf{Complement} \\
\midrule
Recorded model calls & 86 & --- \\
Carried token-usage data & 77 & 9 predate the usage field \\
Carried a measured cost & 67 & 19 predate the cost field \\
Token counts estimated & 10 & 76 measured (char-count proxy) \\
\bottomrule
\end{tabular}
\end{table}

\begin{figure}[t]
  \centering
  \includegraphics[width=0.92\textwidth]{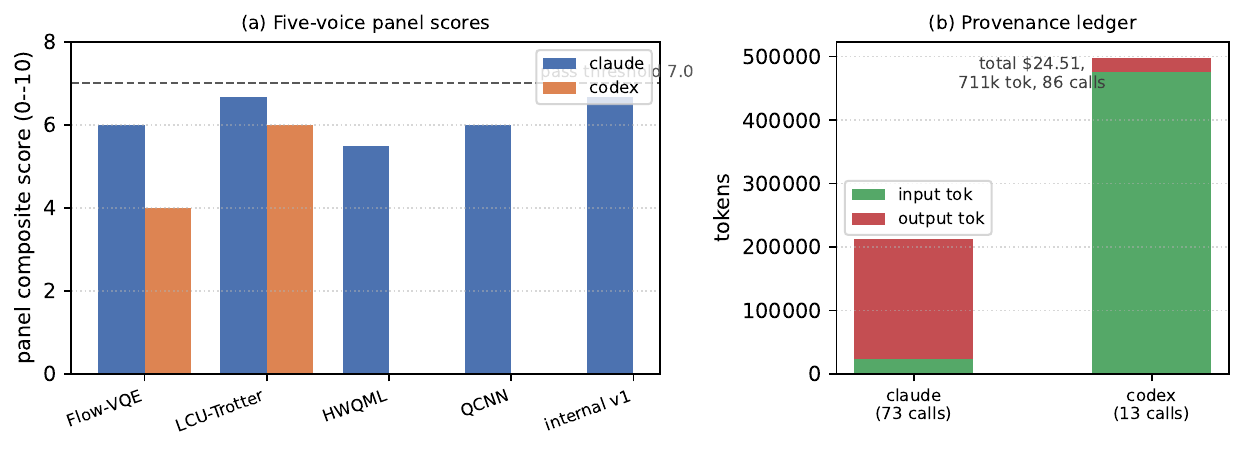}
  \caption{Deployment results from real run artifacts. (a) Five-voice panel
  composite scores per manuscript; no external paper cleared the $7.0$ pass
  threshold, and the two papers run under both vendors show the cross-vendor
  behavior discussed in the text (LCU-Trotter near-agreement, Flow-VQE a
  reject/major-revisions split). (b) The provenance ledger: input and output
  tokens by vendor across $86$ recorded calls, totalling $24.51$ US dollars and
  $711$k tokens. Codex cost was not recorded by the backend. The Majorana
  manuscript, recovered from archived sidecars after this figure was generated,
  is listed in Table~\ref{tab:panel} but omitted from panel~(a).}
  \label{fig:results}
\end{figure}

A Claude backend served seventy-three calls and a Codex backend served thirteen. Cost was recorded for the Claude calls except for six older Claude sidecars that predate the cost field, which is why the measured-cost count in Table~\ref{tab:sidecars} is sixty-seven rather than seventy-three; no Codex call carried a cost. Of the sixty-seven sidecars that also recorded a model identifier, three tiers appear, an Opus-class model on thirty-four calls, a Haiku-class model on twenty-nine, and a second Opus snapshot on four (34+29+4=67), confirming that the framework exercised more than one model tier; the remaining nineteen sidecars predate the model-identifier field.

\subsection{Panel outcomes}

The reviewer stage simulates a five-voice panel consisting of a physics scrutineer, a novelty assessor, an evidence auditor, a devil's advocate, and an editor-in-chief, with a passing threshold of seven out of ten. No external manuscript crossed the threshold. Table~\ref{tab:panel} lists the recorded scores and Fig.~\ref{fig:results}(a) plots them against the cost ledger of Fig.~\ref{fig:results}(b). These verdicts are not calibrated against human referee scores; Section~\ref{sec:experiments} compares them with the publication outcomes that four of the five external manuscripts have since acquired, and beyond that coarse check they should be read as the framework's raw output, not as accuracy estimates.

\begin{table}[H]
\centering
\caption{Recorded five-voice panel scores and verdicts. The composite is the mean of the numeric voice scores; the devil's advocate and editor-in-chief issue categorical verdicts without numeric scores. Because those two categorical verdicts also feed the recommendation, identical composites can map to different verdicts (e.g.\ $6.67$ yields ``minor revisions'' for LCU-Trotter but ``major revisions'' for the internal manuscript); the verdict is therefore not a function of the composite alone. The Majorana row was recovered from archived sidecars after Fig.~\ref{fig:results}(a) was generated and appears in this table only.}
\label{tab:panel}
\footnotesize
\begin{tabular}{l l c l}
\toprule
\textbf{Manuscript} & \textbf{Backend} & \textbf{Score} & \textbf{Verdict} \\
\midrule
Flow-VQE & Codex & 4.0 & reject \\
Flow-VQE & Claude & 6.0 & major revisions \\
LCU-Trotter & Codex & 6.0 & major revisions \\
LCU-Trotter & Claude & 6.67 & minor revisions \\
HWQML & Claude & 5.5 & major revisions \\
QCNN & Claude & 6.0 & major revisions \\
Majorana & Claude & 6.33 & major revisions \\
internal v1 & Haiku/Opus & 6.67 & major revisions \\
\bottomrule
\end{tabular}
\end{table}

\subsection{Flow-VQE: a contested manuscript}

\begin{figure}[t]
  \centering
  \fbox{\includegraphics[page=1,width=0.95\columnwidth]{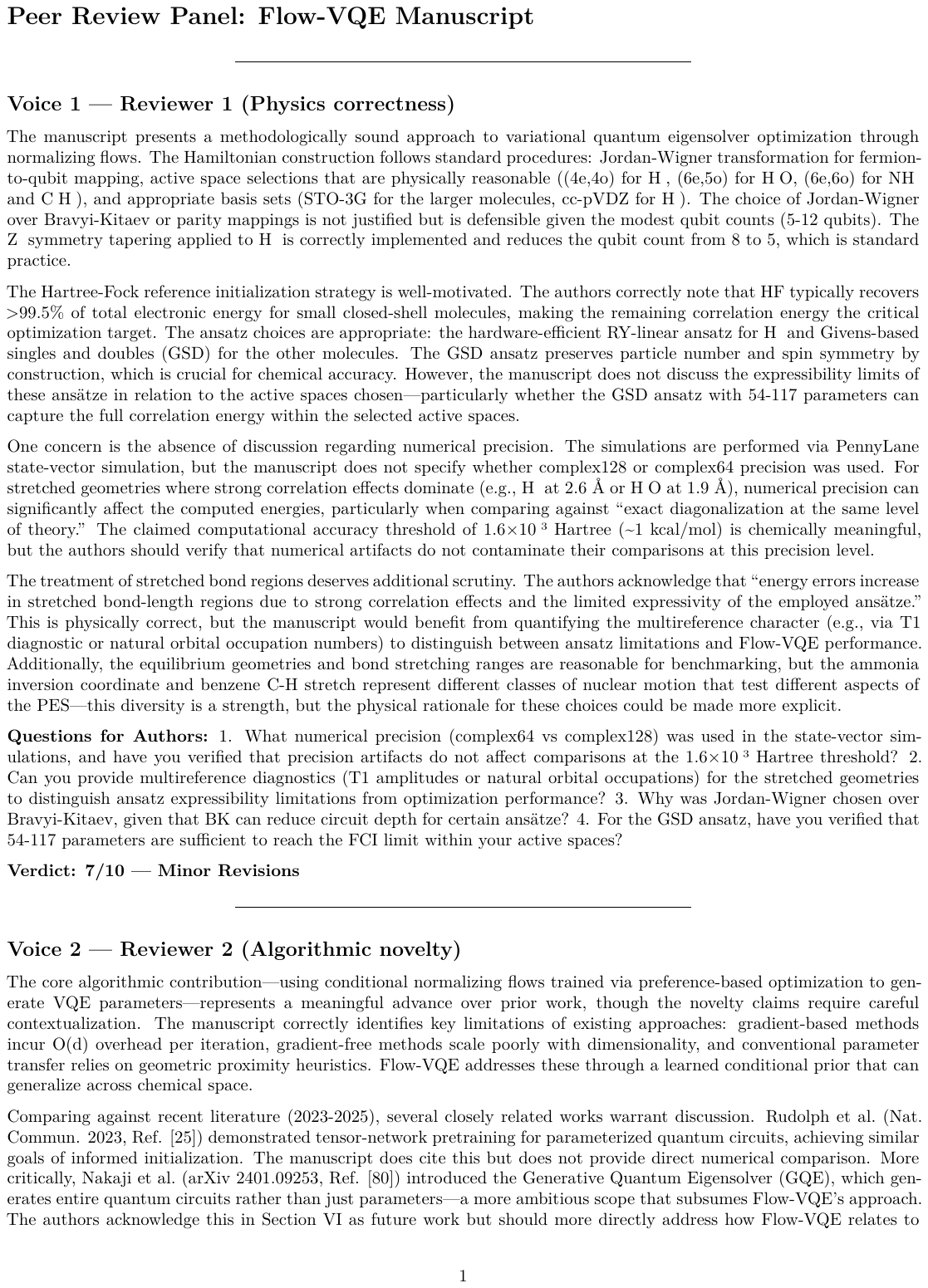}}
  \caption{Real framework output (exhibit): the first page of the actual
  five-voice review panel for Flow-VQE, taken verbatim from the QuantumNovelty
  repository (\texttt{examples/paper\_reviews/flowvqe/02\_reviewer\_panel/}).
  The per-voice verdicts and the quoted critiques in the text are read from this
  artifact.}
  \label{fig:exhibit-review}
\end{figure}

Flow-VQE adapts preference-based training to variational quantum eigensolvers~\cite{zou2025flowvqe, peruzzo2014vqe}. The panel found a genuine methodological idea wrapped in an overstated empirical claim. The novelty assessor flagged baseline inflation at the headline level, writing that ``the `two orders of magnitude' headline claim requires comparing against raw gradient descent without momentum, which is a strawman'' and that the baseline optimizers ``were run with identical learning rates rather than individually tuned.'' The evidence auditor made statistical rigor a barrier, noting that ``CNOT counts and energy errors are point estimates without confidence intervals or multi-seed variance'' and that four-significant-figure energies ``without standard deviations are unverifiable.'' The devil's advocate was blunter, writing that ``this manuscript exemplifies a troubling trend in quantum computing: impressive-sounding improvements over carefully chosen baselines that dissolve under scrutiny,'' and pointing to a hidden failure where ``Figure 2(b) shows Flow-VQE-S losing to Adam, mentioned in passing but not analyzed.'' The editor-in-chief nonetheless credited the core idea, calling ``the preference-based training adapted from reinforcement learning a genuine methodological contribution, even if the components are individually known.''

The two vendors disagreed across the decision boundary. The Codex panel reached a composite of 4.0 with a reject, while the Claude panel held at 6.0 with major revisions. The divergence is traceable in the artifacts to the cost-accounting stance. The Codex panel's required-actions list demanded recomputing ``all headline claims with complete cost accounting, including pretraining, generated-candidate evaluation, fine-tuning, shot assumptions, and all failed runs,'' while the Claude panel stopped at reframing the headline comparison. This is the cross-vendor falsifiability principle observed in practice, although as Section~\ref{sec:limitations} notes it arose from independent per-vendor runs rather than the paired primitive. For verifiability we state the full voice breakdown behind the two composites, read from each run's quality-gate sidecar. The Claude panel scored 7 (physics scrutineer, minor revisions), 6 (novelty assessor, major revisions), and 5 (evidence auditor, major revisions), with the devil's advocate (reject) and the editor-in-chief (major revisions) issuing categorical verdicts without numeric scores; the composite 6.0 is the mean of the three numeric voices. The Codex panel scored 5, 4, and 3 on the same three voices with categorical rejects from the remaining two, giving 4.0. The exhibited Voice-1 score of 7/10 in Fig.~\ref{fig:exhibit-review} is therefore consistent with, and now checkable against, the composite it enters.

\subsection{LCU-Trotter: near-agreement and five fallacies}

LCU-Trotter is a theory paper combining linear-combination-of-unitaries with Trotter compensation~\cite{zeng2025lcutrotter}. The two vendors nearly agreed, at 6.0 and 6.67, because the formal results were found sound and the disagreement concerned comparison methodology rather than fundamentals. Reviewer three objected that ``the y-axis label conflates different gate definitions'' between deterministic Trotter rotations and the random Pauli rotations of the compensation scheme, ``whose classical angle-computation overhead is not accounted for.'' The fallacy stage emitted five named findings, all at medium severity, listed in Table~\ref{tab:fallacies}. They illustrate the quantum-specific taxonomy, from an asymptotic-only claim that rests on big-$O$ formulas ``with no numerical constants or crossover points'' to a conflated-regimes finding that a two-orders-of-magnitude advantage ``mixes loose analytical bounds in one panel with tight commutator-aware bounds in another.'' Most of the taxonomy (cherry-picked baseline, hasty generalization) is domain-agnostic, but at least one category is genuinely quantum-specific: \emph{active-space-handwave}, fired on Flow-VQE (Table~\ref{tab:fallacies}), targets the quantum-chemistry move of claiming broad resource savings while the reported experiments reach only a small active space, a failure mode that has no analogue outside electronic-structure simulation. A second, \emph{ad-hoc-precision-floor}, targets accuracy claims that ignore the sampling overhead specific to linear-combination-of-unitaries circuits. We nonetheless treat the taxonomy's quantum-specificity as only partially demonstrated, since most emitted instances were domain-agnostic.

\begin{table}[t!]
\centering
\caption{Named fallacies emitted by the fallacy stage, with abbreviated real instances. Two of the seven categories (active-space-handwave and ad-hoc-precision-floor) require quantum-domain knowledge to identify; the remaining five are general scientific-reasoning failures.}
\label{tab:fallacies}
\footnotesize
\begin{tabular}{l l p{0.55\textwidth}}
\toprule
\textbf{Fallacy} & \textbf{Manuscript} & \textbf{Real instance (abbreviated)} \\
\midrule
cherry-picked-baseline & LCU-Trotter & Competitor shown as order $n^2$ while ignoring commutator-aware improvements \\
asymptotic-only-claim & LCU-Trotter & Superiority rests on big-$O$ formulas with no constants or crossover points \\
conflated-regimes & LCU-Trotter & Two-orders claim mixes loose and tight bounds across figure panels \\
hasty-generalization & LCU-Trotter & Universal recommendations from only two Hamiltonian families \\
ad-hoc-precision-floor & LCU-Trotter & Accuracy gains assume perfect gates; sampling overhead omitted \\
active-space-handwave & Flow-VQE & Broad savings claimed while experiments reach only twelve-qubit active spaces \\
circular-reasoning & HWQML & Full-rank quantum Fisher matrix used as evidence of arbitrary-state reachability \\
\bottomrule
\end{tabular}
\end{table}

\subsection{HWQML and QCNN}

The hardware-efficient quantum machine learning manuscript scored 5.5 with major revisions. The novelty assessor flagged a load-bearing unproved conjecture and required the authors to ``either prove the conjecture or substantially expand numerical evidence to larger qubit counts.'' The quantum convolutional neural network manuscript scored 6.0, and reviewer three identified an unfalsifiable narrative, writing that ``if the classical surrogate succeeds the dataset is declared locally easy, and if the network fails the dataset is too entangled,'' together with a reporting concern that a table ``presents best out of five runs, which cherry-picks rather than reporting mean performance.''

\section{Patent Examination Case Study}
\label{sec:patent}

\subsection{Background: how a US patent application is prosecuted}

A US patent application pairs a \emph{specification} (the written description and
drawings) with numbered \emph{claims} that define the legal scope sought; the
claims are the operative text and the specification only supports them. An
\emph{independent} claim recites every element it requires, and a
\emph{dependent} claim narrows a parent by adding limitations. Validity is judged
claim by claim.

After filing, a USPTO examiner tests each claim against four statutes:
\S101 eligibility, the \emph{Alice}~\cite{alice2014} abstract-idea bar that
quantum-algorithm claims must clear; \S102 anticipation, where a single reference
discloses every element; \S103 obviousness, a combination of references a skilled
person would make, per \emph{KSR}~\cite{ksr2007}; and \S112, enablement, written
description, and \emph{definiteness} (a common defect being a missing
\emph{antecedent basis}, a claim saying ``the X'' with no earlier ``an X''). The
examiner issues an \emph{Office Action} listing rejections and cited art; the
applicant replies by \emph{amending} the claims or \emph{arguing} the rejections
are wrong, and prosecution iterates until allowance or final rejection.
QuantumNovelty instantiates both sides: its six-voice examiner panel produces the
Office Action, and its drafting skill produces the application.

\subsection{The examiner panel and a worked Office Action}

The patent subsystem applies the same agent paradigm to United States patent law. The patent-reviewer skill runs a six-voice examiner panel. A primary examiner handles section 101 eligibility and the overall disposition, with the abstract-idea inquiry following the \emph{Alice} framework~\cite{alice2014}, a section 102 examiner handles anticipation and names a single anticipatory reference, a section 103 examiner builds obviousness combinations under \emph{KSR}~\cite{ksr2007}, a section 112 examiner handles enablement, written description, and definiteness, a quantum technical specialist checks physical operability against the no-cloning theorem, the Holevo bound, and fault-tolerance overreach, and a supervisory examiner synthesizes the disposition.

\begin{figure}[t]
  \centering
  \fbox{\includegraphics[page=1,width=0.95\columnwidth]{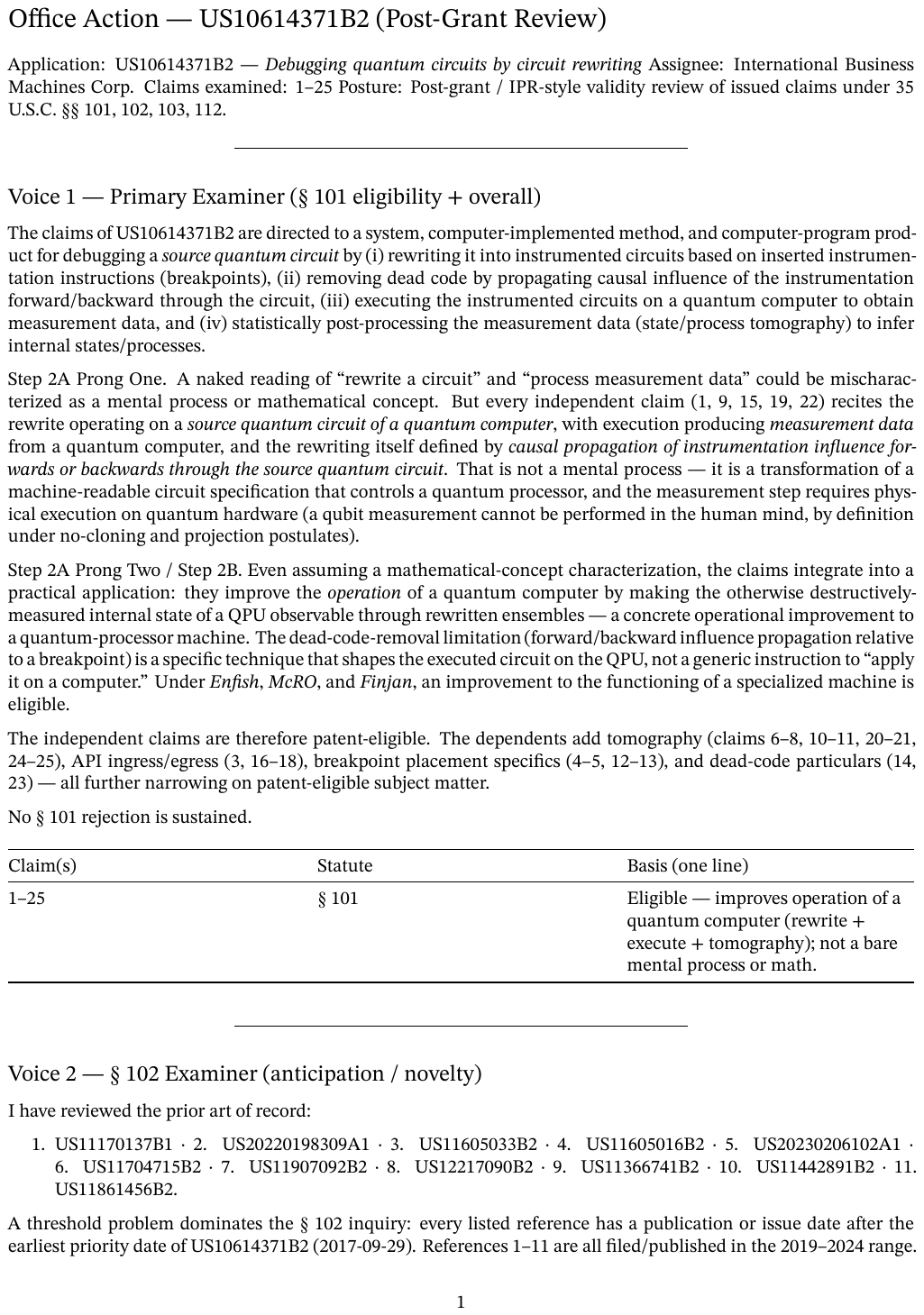}}
  \caption{Real framework output (exhibit): the first page of the actual
  six-voice examiner Office Action for granted patent US10614371B2, taken
  verbatim from the QuantumNovelty repository
  (\texttt{datasets/quantum\_patent\_office\_actions/eval\_fixed2/US10614371B2/}).
  The analysis quoted below is read from this artifact.}
  \label{fig:exhibit-oa}
\end{figure}

We ran the panel on US10614371B2, \emph{Debugging quantum circuits by circuit rewriting} (International Business Machines; filed 2017-09-29, granted 2020-04-07), a quantum-software patent chosen because its prosecution outcome is known: the claims were allowed after a single non-final Office Action. The panel run (one Opus-class call, 106.5 seconds, 1.20 US dollars, all recorded in the run sidecar) examined all twenty-five claims. The section 101 examiner sustained no eligibility rejection, finding the claims directed to a concrete debugging apparatus rather than an abstract idea. The section 102 and 103 examiners were seeded with the eleven references listed on the patent's public citation record, and both declined to use any of them, for a reason a human examiner would also have to honor: every seeded reference post-dates the patent's 2017-09-29 priority date. The section 103 examiner wrote that ``as a matter of statutory threshold, they cannot be used as \S~103 prior art against this patent on the record before me.'' Declining temporally ineligible art, rather than force-fitting an obviousness combination from whatever references were offered, is the conservative behavior one would want here; we note it is also the trivially correct reading of the priority dates, not by itself evidence that the falsification layer generalizes. We stress that all panel outputs remain \emph{model-proposed candidate} analyses, not adjudicated findings.

The supervisory examiner synthesized an allowance recommendation at a confidence of nine out of ten, which agrees with the real-world disposition: the patent was granted. That agreement is one data point, not a calibration, and Section~\ref{sec:experiments} reports the panel's behavior across fourteen granted patents, where the picture is far less flattering. The run also exposed a genuine defect in our own tooling, which we report rather than suppress: the deterministic parser that converts panel prose into the structured Office Action record found no canonical disposition block and fell back to the per-voice rejection tables, recording a twenty-five-claim section 101 rejection that the prose of every voice contradicts. The structured record and the supervisory synthesis therefore disagree inside the archived artifact. Root-causing this defect exposed a two-cell table-parsing pattern that could not see the basis column in which a voice wrote ``Eligible,'' so allowances could silently flip to rejections in the structured record; the repaired parser treats the supervisory disposition as authoritative when the canonical block is absent, preserves the per-voice tables in a separate field, and sets an explicit \texttt{parse\_conflict} flag. The repair is exercised by the calibration sweep of Section~\ref{sec:experiments}, where the flag fired on four of nine runs and no silent flip survived.

In drafting mode the patent subsystem inverts this analysis. From an invention disclosure it produces a full filing package containing a title, abstract, background, summary, drawings plan, detailed description, examples, claims, claim-compliance notes, application formalities, and a prosecution checklist, together with a deterministic run manifest and the provenance marker. Examination and drafting therefore share one statutory model, used once to attack a set of claims and once to construct one.

\section{Validation Experiments}
\label{sec:experiments}

The case studies above report what the framework produced on its deployment corpus. To probe specific mechanisms more directly, and in response to the gaps a hostile reader would attack first, we ran five small additional experiments. We report them with their negative results intact.

\subsection{The novelty audit, demonstrated end to end}

The strict-domination novelty audit is the framework's headline mechanism, so we exercise it on real input rather than only describing it. We drove the ansatz-discovery loop to a small archive of candidate circuits for a two-qubit Hamiltonian and scored each with the bundled state-vector evaluator (seed 42, 400 SPSA iterations). In the first run, five candidate circuits form a genuine Pareto front: the audit returns \textsc{interpolation} for four and \textsc{rediscovery} for the Bell-state circuit, and certifies none as novel, a verdict consistent with the strict-domination criterion, since no circuit dominates the others on all axes; ``correct'' in a calibrated sense would require external ground truth this experiment does not have. In the second run, a single candidate (C3-MinZZ; energy error $1.7\times10^{-3}$ Ha, four operations, one CNOT) is compared against three baselines deliberately constructed to lie strictly above it on every axis; the audit returns \textsc{strict-domination}, exercising that code path. We disclose plainly that the second run's baselines are notional values chosen to be strictly worse, not numbers extracted from the cited references, so this run demonstrates the mechanism rather than a research finding. A genuine domination claim requires a baseline catalog populated from the literature, which connects directly to the retrieval experiment below.

\subsection{Adversarial validation of the audit gates}
\label{sec:adversarial}

The demonstrations above run the audit on a real Pareto front and on notional dominators; a hostile reader will still ask whether the deterministic gates actually catch an overclaim they were built to catch, because no manuscript in the deployment corpus ever drove the strict-domination or recompute gate against a genuine violation. We therefore constructed a planted adversarial corpus in which the ground truth is fixed by construction (seed 42; \path{adversarial_audit_20260706.json}), and we report each rate with a Wilson $95\%$ interval so the small sample sizes are not read as more than they are. For the strict-Pareto-domination gate we built eight candidate circuits \emph{presented} as novel strict-dominators that are in fact exact ties, ties within the floating-point tolerance, single-axis trade-offs, dominated points, or (in one edge case) a novelty claim made against an \emph{empty} baseline catalog, checked against a three-baseline catalog augmented with eight genuine dominators; the gate rejected all eight ($8/8$; Wilson $95\%$ $[0.68,1.0]$), returning \textsc{rediscovery}, \textsc{dominated}, or \textsc{interpolation}, while certifying all eight genuine dominators ($8/8$; $[0.68,1.0]$) as \textsc{strict-domination}. For the numerical-recompute gate we planted seven ratio claims whose displayed value contradicts the recomputed quotient (for example ``$700/50 = 3.0\times$'', which recomputes to $14.0$) alongside seven internally consistent ratios, including cases where the displayed value is a legitimate rounding of the quotient; the gate flagged all seven fabricated ratios ($7/7$; $[0.65,1.0]$) and raised no false alarm on the seven correct ones ($0/7$; $[0.0,0.35]$). On this planted set the two deterministic gates achieve perfect separation, and the intervals confirm the point estimates are bounded away from chance despite the small $n$. This is a test of the gate logic, not of end-to-end performance on real manuscripts, where the hard part is extracting the right numbers and baselines rather than comparing them once extracted; it establishes that the mechanisms do what the paper claims once correct inputs reach them, a property the deployment corpus never exercised on a genuine violation. Table~\ref{tab:adversarial} summarizes the four rates with their intervals.

\begin{table}[t]
\centering
\caption{Adversarial validation of the two deterministic gates on a planted corpus with known ground truth (seed 42; \texttt{adversarial\_audit\_20260706.json}). Each row is a gate--condition pair; $k/n$ counts correctly-handled cases over planted cases, and the Wilson column is the $95\%$ score interval for that rate. The gates separate planted overclaims from genuine claims with no false positives; the intervals are wide only because $n$ is small by construction, not because any case was mishandled.}
\label{tab:adversarial}
\small
\setlength{\tabcolsep}{6pt}
\begin{tabular}{@{}llcc@{}}
\toprule
\textbf{Gate / outcome} & \textbf{Planted condition} & \textbf{$k/n$} & \textbf{Wilson $95\%$} \\
\midrule
\multicolumn{4}{@{}l@{}}{\emph{Strict Pareto-domination}} \\
\quad overclaims caught & ties, trade-offs, dominated, empty catalog & $8/8$ & $[0.68,\,1.00]$ \\
\quad genuine certified & true strict dominators & $8/8$ & $[0.68,\,1.00]$ \\
\midrule
\multicolumn{4}{@{}l@{}}{\emph{Numerical recomputation}} \\
\quad fabricated caught & displayed $\neq$ recomputed quotient & $7/7$ & $[0.65,\,1.00]$ \\
\quad false positives & correct ratios (incl.\ rounding) & $0/7$ & $[0.00,\,0.35]$ \\
\bottomrule
\end{tabular}
\end{table}

\subsection{What the framework's claims rest on}
\label{sec:ledger}

Because this paper makes a deliberately narrow claim, Table~\ref{tab:ledger} states each checkable claim against the artifact that supports it, the denominator it is computed over, and the caveat that bounds it. No claim in the table depends on human labels, and no claim asserts agreement with a human referee's judgment; each is a property of the framework's own recorded outputs or of a planted corpus with known ground truth.

\begin{table}[t]
\centering
\caption{The paper's checkable claims and what each rests on. Every entry is verifiable from an on-disk artifact without human annotation; none claims agreement with a human referee.}
\label{tab:ledger}
\footnotesize
\begin{tabular}{p{0.24\textwidth} p{0.20\textwidth} p{0.14\textwidth} p{0.32\textwidth}}
\toprule
\textbf{Claim} & \textbf{Artifact} & \textbf{Denominator} & \textbf{Result and caveat} \\
\midrule
Deterministic gates separate planted overclaims from genuine claims & \path{adversarial_audit_20260706.json} & 8 Pareto + 14 ratio cases & $8/8$ overclaims caught, $8/8$ genuine certified, $7/7$ fabricated ratios flagged, $0/7$ false positives (Wilson $95\%$ intervals in text). Tests gate logic, not extraction of correct inputs. \\
Panels are directionally more conservative than the public outcome & Table~\ref{tab:panel}; grant dispositions & 4 accepted papers; 14 granted patents & Panel scored $4.0$--$6.67$ (bar $7.0$) on four accepted papers; patent layer over-rejected $14/14$ granted patents before repairs, and a post-repair re-run allowed $5/9$ (\S\ref{sec:experiments}). One-sided sample: fixes the sign of the bias, not its magnitude. \\
Every recorded model call writes a provenance sidecar & 86 per-call sidecars & 86 recorded calls & 86 sidecars recorded (denominator is recorded calls, not a separately-audited call count); 77 carry token-usage data and 67 carry measured monetary cost (full partition, including the measured-vs-estimated split, in Table~\ref{tab:sidecars}). Codex backend logged no cost; $25/28$ timestamped run dirs are empty placeholders. \\
\bottomrule
\end{tabular}
\end{table}

\subsection{Cost-quality-evidence, by dimension}

All five external papers in the deployment corpus received an identical CQE composite of twenty-three. Table~\ref{tab:cqebreak} gives the per-dimension breakdown that explains it. The geometric mean is dragged down by two dimensions: novelty rigor ($8$) and reproducibility ($20$). Crucially, the novelty-rigor probe records ``\texttt{novelty\_verdict.json not found}'', ``baseline catalog has 0 rows'', and ``archive rows: 0'' for these runs, so the score of $8$ measures the \emph{absence of novelty-audit artifacts} (these were front-half-only review runs) rather than a judgment that the papers lack novelty. The constant composite is thus an honest signal that the full pipeline was not run on these papers, not a degenerate metric; a run that populates the archive scores higher on those axes by construction. The operational conclusion, however, must be stated plainly: in this deployment the composite functioned as a pipeline-completeness indicator, not a quality discriminator, and a constant composite across papers carries no information about their relative quality. We therefore label it accordingly here and in Table~\ref{tab:cqebreak}, and we do not use the CQE to rank the reviewed papers anywhere in this article.

\begin{table}[t]
\centering
\caption{Per-dimension CQE scores for the five external papers (from each run's \texttt{cqe\_scores.json}). The composite is the geometric mean; novelty rigor and reproducibility dominate the result. In this deployment the constant composite functions as a pipeline-completeness indicator rather than a quality discriminator.}
\label{tab:cqebreak}
\footnotesize
\begin{tabular}{l c c c c c c c}
\toprule
\textbf{Paper} & \textbf{Nov} & \textbf{Rep} & \textbf{Meth} & \textbf{Fals} & \textbf{Dom} & \textbf{Comm} & \textbf{Comp} \\
\midrule
Flow-VQE & 8 & 20 & 27 & 30 & 30 & 40 & 23 \\
HWQML & 8 & 20 & 27 & 30 & 30 & 40 & 23 \\
LCU-Trotter & 8 & 20 & 27 & 30 & 30 & 40 & 23 \\
Majorana & 8 & 20 & 27 & 30 & 30 & 40 & 23 \\
QCNN & 8 & 20 & 27 & 30 & 30 & 40 & 23 \\
\bottomrule
\end{tabular}
\end{table}

\subsection{Literature-retrieval recall}

Because the novelty audit is only as good as the baseline catalog its literature agent supplies, we measured retrieval recall directly. We ran \texttt{literature\_surfacer} (Crossref, arXiv, Semantic Scholar; forty deduplicated cards per topic) on three topics with an unambiguous seminal reference and checked whether that reference was returned. The result is a candid weakness: recall was one of three. The barren-plateaus query returned McClean \emph{et al.} (2018) as its top card, but the variational-quantum-eigensolver query did \emph{not} return Peruzzo \emph{et al.} (2014), and the quantum-convolutional-network query did \emph{not} return Cong, Choi, and Lukin (2019); the latter query also drew eleven of forty off-topic cards, because the Semantic Scholar source over-retrieved on ``convolutional neural networks'' without enforcing the quantum constraint. A novelty audit fed by this retrieval would miss the defining prior art for two of three topics. Improving retrieval precision and recall is therefore a prerequisite for the novelty audit to be trustworthy on real papers, and is the single most actionable engineering gap the framework has. Until recall on such probes clears at least two of three seminal references, novelty-audit verdicts on real papers should be treated as indicative rather than authoritative.

\subsection{Reviewer run-to-run stability}

Because the panels are language-model outputs, their reproducibility is a fair concern. As a spot check rather than a variance estimate ($n=2$ runs on one manuscript), we ran the reviewer twice, independently, on the same manuscript (the hardware-efficient quantum machine learning paper) in quick mode. Both runs returned the same recommendation category, \emph{minor revisions}, so the verdict was stable across runs at the categorical level on this single spot check. Quick mode emits a recommendation rather than a numeric gate, so we did not obtain a composite-score variance; measuring score-level variance across repeated full-panel runs remains future work, and the deterministic gates of Section~\ref{sec:audit} are by construction invariant across runs.

\subsection{Referee-panel calibration against publication outcomes}

After the deployment's runs were archived, four of the five external manuscripts acquired a public editorial outcome, and all four were accepted: LCU-Trotter as PRX Quantum \textbf{6}, 010359; Flow-VQE in npj Quantum Information~\cite{zou2025flowvqe}; the hardware-efficient machine-learning manuscript in \emph{Quantum}; and the quantum-convolutional-network manuscript in PRX Quantum. The panel's six runs on these four papers scored between 4.0 and 6.67, and none crossed the 7.0 pass bar (Table~\ref{tab:panel}). At the outcome level, then, the referee panel is directionally more conservative than the venues that accepted the same work, on a one-sided sample (all four outcomes are acceptances), which is the same directional finding as the patent calibration below. Because the sample contains no rejected papers, this indicates the sign of the panel's bias, not its magnitude, and a panel that rejected everything would score identically here. Three caveats bound the comparison. Acceptance is a coarse label mediated by revision, and the panel reviewed arXiv versions that may predate the accepted ones; venue thresholds differ from the panel's fixed bar; and four papers with a single outcome class support no precision or recall estimate, for which a corpus containing rejected papers is still required. What the comparison does establish, cheaply and from the public record, is the direction of the panel's bias: like the per-voice patent layer, it under-accepts relative to real editorial outcomes.

\subsection{Patent-panel calibration against granted patents}

Because the panel's judgments need an external standard, we assembled a public-record dataset of twenty-eight granted US quantum-computing patents (metadata, claims text, and examiner-cited references; the full Office Action texts were not programmatically retrievable, since the USPTO text endpoints reject automated clients, and manual retrieval through Patent Center remains open work). Every patent in the set was granted, so the panel's disposition can be scored against a known outcome. An evaluation sweep over fourteen of them, one per unique assignee (the worked example US10614371B2 of Section~\ref{sec:patent} is one of these fourteen), is bluntly negative for the per-voice machinery: the panel issued a rejection-bearing Office Action for fourteen of fourteen granted patents, a one-hundred-percent over-rejection rate at the per-voice level, rejecting 17.6 claims per patent on average, with section 103 raised against twelve patents and section 112 against thirteen. Prior-art overlap between the panel's independently proposed references and the examiner-cited lists was zero across all fourteen. Three design caveats bound these numbers. The panel reviewed the final granted claims, a harder target than the as-filed claims the real examiners saw; the sweep did not feed the specification to the panel, so the section 112 counts partly measure missing input rather than claim defects; and the examiner-cited reference lists come from the public citation record rather than from the Office Action documents themselves. The sweep did its job: it exposed three concrete defects, and repairing them is itself a result. The audit traced the over-rejection to (i) the two-cell parsing defect of Section~\ref{sec:patent}, which silently converted allowances into rejections in the structured record; (ii) reference seeding that had scraped \emph{forward} citations (papers citing the patent) instead of backward ones, offering the panel prior art that post-dates the priority date, now blocked by a deterministic temporal guard; and (iii) the missing specification input. After these repairs, a re-run over nine granted patents (five assignees) recovers allowance on five of nine, cuts mean rejected claims from 17.6 to 1.8 per patent, raises no section 102 or 103 rejection at all, and confines the remaining over-rejection to narrow section 112 findings on two to six claims per patent; the new parse-conflict flag fired on four of nine runs, each resolved to the supervisory disposition rather than silently flipped. The two sweeps differ in sample and inputs, so the improvement confounds the three repairs, and a per-repair ablation with real Office Action texts is the next patent-mode experiment. These are the calibration numbers this paper can offer, and they replace an untestable resemblance claim with a testable, improvable one: on this one-sided sample of granted patents the per-voice layer began markedly more rejection-prone than the known outcomes support, and deterministic repairs, found by calibrating against those public outcomes, moved it toward the record (though, as above, an all-granted sample fixes the direction of the bias, not its magnitude).

\section{Implementation Validation}
\label{sec:validation}

\paragraph{Platform.} For reproducibility we state the platform at the level a re-runner needs. QuantumNovelty version 1.0.0 is approximately 14{,}800 lines of Python and 1{,}500 lines of shell, requires Python 3.11 or later, and is released under the MIT license. The twenty-two skills run on the five-backend interface of Section~\ref{sec:arch}; the default path shells to the Claude Code command-line interface and requires no API key. The deployment corpus ran against three pinned model identifiers, an Opus-class model on thirty-four calls, a Haiku-class model on twenty-nine, and a second Opus snapshot on four, with each call's identifier recorded in its sidecar. All Python executes through a shared virtual environment so that a skill and any subprocess it spawns resolve to the same interpreter and scientific stack, which keeps a result reproducible regardless of which pipeline produced it.

\paragraph{Test suite.} Separate from the deployment, the framework carries a test suite that collects 183 cases across four modules. The collected count expands from 110 test functions (32+23+35+20) to 183 collected cases because the smoke and telemetry functions are parametrized once per matching repository file, so a single function contributes as many cases as there are files it sweeps. Only one of the four modules invokes a live model. A smoke module (32 functions) performs pure command-line and static-analysis assertions with no model call, parametrized over every Python and shell file in the repository. A multi-mode module (23 functions) tests the natural-language dispatcher's routing from a free-text request to the correct pipeline, skill, and mode, also without a live call. A telemetry module (35 functions) checks the aggregation and ledger logic that underpins the provenance guarantees, again model-free. Only the real-skills module (20 functions) exercises live skill invocations against a backend. The bulk of the suite is therefore deterministic and runnable offline. Two zero-model introspection commands provide a second check, since listing the skills and printing the stage table both run without a model call and confirm that every pipeline references only skills that exist, and that every skill is discoverable, which makes both commands safe to run in continuous integration.

\section{Discussion and Limitations}
\label{sec:limitations}

The deployment answers what the framework computes and what it costs. It does not answer whether the framework is correct, and the gaps are worth stating plainly.

There is no per-review human baseline. No artifact in the corpus contains human referee reports or score-level judgments, so fine-grained agreement between the framework and human experts cannot be computed. The coarse outcome-level standards that do exist in the public record, journal acceptance for four of the five external manuscripts and grant dispositions for the patent calibration set, are compared in Section~\ref{sec:experiments}; both comparisons show the framework more rejection-prone than the human record, but neither supports a precision or recall estimate. Every quantitative statement in Sections~\ref{sec:review} and~\ref{sec:patent} is a statement about the framework's own outputs. In particular, the named fallacies of Table~\ref{tab:fallacies} are model-emitted findings whose precision (the fraction that a human would confirm as genuine) has not been adjudicated; we report them as produced, not as verified, and a true-versus-false human pass over each finding is the natural next measurement.

Generation mode is only partially exercised. The deployment corpus ran the review and audit half of the framework, and no generated full paper draft was produced there. The strict-domination novelty audit, the framework's headline mechanism, is now demonstrated end to end in Section~\ref{sec:experiments}: it certifies none of a real five-circuit Pareto front as novel, consistent with the domination criterion, and returns strict-domination on a constructed dominating case. That demonstration is small and used notional baselines; the adversarial validation of Section~\ref{sec:adversarial} separately confirms the gate logic catches planted overclaims (eight of eight false-novelty circuits, seven of seven fabricated ratios) with no false positives. The remaining gap is running the mechanism against a literature-populated baseline catalog at scale on real inputs, where the hard part is extracting correct baselines rather than comparing them. Closing that gap depends on the retrieval improvements of Section~\ref{sec:experiments}.

The cross-model falsifiability guard was not exercised as a paired primitive in this corpus. The independent per-vendor runs of Section~\ref{sec:review} are a weaker substitute, and the framework's own falsifiability probe correctly recorded an empty vendor set for the affected runs. A genuine multi-vendor consensus claim would require new runs of the paired primitive.

The cost-quality-evidence composite is keyword-based rather than expert-graded. Its per-dimension breakdown (Table~\ref{tab:cqebreak}) shows the identical composite of twenty-three is driven by the novelty-rigor and reproducibility probes, which register the absence of novelty-audit artifacts rather than a judgment of paper quality; its correlation with real quality is untested. We did not calibrate the panels or the composite against known outcomes, because the framework's calibration mode requires a labeled gold set of accepted and flawed papers (it exits without one) that we did not assemble; this is a clear item of future work. Separately, the Codex backend did not record cost, and twenty-five of the twenty-eight timestamped run directories are empty placeholders. Finally, literature retrieval recall is currently weak (one of three seminal references on a small probe, Section~\ref{sec:experiments}), so the novelty audit's baseline catalog cannot yet be trusted to be complete; improving retrieval precision and recall, assembling a calibration gold set, measuring score-level reviewer variance across repeated full-panel runs (categorical stability is shown in Section~\ref{sec:experiments}), and retrieving the issued Office Action texts to deepen the granted-patent calibration of Section~\ref{sec:experiments} together form the validation plan this paper does not yet complete.

Finally, the panels are simulations and the framework is decision-support infrastructure. A simulated five-voice referee panel is not editorial peer review and a simulated examiner panel is not a USPTO action. The value of the framework is that it produces these artifacts in a uniform, recomputed, costed, and reproducible form, not that it carries editorial or legal authority. The framework is also scoped to quantum computing, since the fallacy taxonomy, the examiner panel, and the baseline catalogs are tuned to that domain.

\subsection{A concrete evaluation protocol}
\label{sec:evalprotocol}

The measurements in this paper are outcome-level (journal acceptance, patent grant) and adversarial (Section~\ref{sec:adversarial}); neither substitutes for agreement with expert referees, which requires human labels this deployment does not have. We therefore specify the protocol that would supply them, so that the gap is a defined experiment rather than an open-ended aspiration, and so that the framework's existing \texttt{calibration} skill (which today exits without a labeled set) has a target format.

\emph{Gold set.} A locked, versioned corpus of 20--30 quantum-computing manuscripts spanning accepted and rejected or withdrawn outcomes, and 10--14 US patent applications spanning granted and rejected or abandoned dispositions, so that both corpora contain both classes and support recall as well as precision. Patents are supplied as as-filed claims with the specification, matching what a real examiner saw, which also fixes the invalid \S~112 input noted above.

\emph{Annotation.} Two to three quantum-computing PhD annotators, blind to the framework's output, independently label each manuscript with an accept/reject recommendation and three to five genuine, specific flaws, and each patent claim set with the prior-art references they would cite. Inter-annotator agreement (Cohen's $\kappa$) is reported to bound the achievable ceiling.

\emph{Metrics.} For review mode: finding-level precision (the fraction of framework findings an annotator confirms genuine), recall (the fraction of annotator-identified flaws the framework surfaces), and their $F_1$; the missed-critical-flaw and hallucinated-criticism rates; verdict agreement with the human recommendation and with the editorial outcome; and score-level stability from three full-panel runs per manuscript (mean $\pm$ SD and categorical agreement rate). For patent mode: claim-level prior-art overlap against the examiner-cited references (overlap@$k$) and the false-rejection rate on the granted subset. These metrics would replace the present directional and adversarial evidence with a calibrated accuracy estimate against human judgment. We specify this protocol rather than execute it: a blind, multi-annotator study at this scale is beyond the resources of this work. Crucially, no claim in this paper depends on it. Our validated claims are deliberately bounded to what is checkable without human labels --- the correctness of the deterministic gates (Section~\ref{sec:adversarial}), the direction of the panels' bias against the public outcome record (Section~\ref{sec:experiments}), and the completeness of the provenance ledger --- and we make no referee-quality or examiner-quality accuracy claim, which is precisely what this protocol would test. We publish the protocol and the gold-set format so that an adopter with access to expert annotators can run it against the framework's existing \texttt{calibration} skill.

\section{Conclusion}
\label{sec:conclusion}

QuantumNovelty applies the skill-orchestrating language agent paradigm to the production and scrutiny of quantum-science claims. It authors papers, drives an ansatz-discovery loop, and drafts patents in generation mode, and it produces referee-style reports and examiner-style patentability-screening reports in review mode, all through composable skills and named pipelines. It constrains certain unfalsifiable claims by construction through a strict Pareto-domination test, numerical recomputation, small-sample confidence reporting, and a cross-model vendor guard, and every run leaves a provenance ledger that records what was computed and at what cost. A first deployment reviewed six quantum manuscripts and a granted US patent at a measured cost of about twenty-four US dollars, producing detailed referee critiques, a partially quantum-specific fallacy taxonomy in action, and a patent-mode calibration against granted patents with known outcomes, while also exposing concretely which of the framework's mechanisms remain to be exercised on real inputs. As agentic systems take on more of the production of quantum results, infrastructure that produces and scrutinizes those results in an auditable form becomes necessary, and QuantumNovelty is a concrete step toward it.

\subsubsection*{Reproducibility Statement}
The framework is open-source under a permissive licence and will be released
publicly on acceptance; an anonymized snapshot is provided as supplementary
material. The adversarial gate-validation of Section~\ref{sec:adversarial} is
fully deterministic: its planted corpus, verdicts, and Wilson intervals are
regenerated by a self-contained script (seed 42) that vendors the same
deterministic gate logic used by the framework's \texttt{novelty\_audit} skill
(no external repository, model call, or absolute path), and the resulting
\path{adversarial_audit_20260706.json} is written beside the script so the
reported rates can be recomputed directly. The run artifacts underlying
Sections~\ref{sec:review}--\ref{sec:experiments} (provenance sidecars, review
outputs, novelty-audit verdicts, and literature-retrieval results) are included
in the anonymized supplement.

\subsubsection*{Broader Impact Statement}
This work builds tooling that produces referee-style and examiner-style
artifacts for quantum papers and patents. The intended use is decision support:
uniform, recomputed, costed, and reproducible artifacts that a human expert
reviews. A misuse risk is that such artifacts are treated as authoritative
verdicts and substituted for human peer review or patent examination; the paper
is explicit that the panels are simulations carrying no editorial or legal
authority, that the framework makes no per-review accuracy claim against human
experts, and that on the public outcome record the panels are directionally
more conservative (more rejection-prone) than the human dispositions. The
deterministic audit gates are designed to constrain, not license, unfalsifiable
claims. We report negative results in full to discourage over-trust.

\subsubsection*{Disclosure of AI use}
The manuscript prose was drafted with large-language-model assistance and then
curated and edited by the authors, who are responsible for every claim. Figures
and tables are not model-fabricated: diagrams are hand-drawn, and all reported
numbers are computed from on-disk run artifacts, each checked against its
artifact before inclusion. Beyond manuscript preparation, this paper carries an
AI-circularity worth flagging: the corpus analyzed in
Sections~\ref{sec:review}--\ref{sec:experiments} was itself produced by the
framework under study (an LLM-backed system), so the study uses an LLM-backed
tool to scrutinize LLM-backed outputs, with no human ground truth
(Section~\ref{sec:limitations}).

\end{document}